\documentclass[aps,10pt, twocolumn, amsmath,amssymb,pra,superscriptaddress,floatfix]{revtex4-2}

\usepackage{graphicx}
\usepackage[dvipsnames]{xcolor}
\usepackage{bm}% bold math

\usepackage{empheq,etoolbox}
\usepackage[makeroom]{cancel}
\usepackage{bbold}
\usepackage{comment}
\usepackage{appendix}

\usepackage{mathabx}
\usepackage{amsmath}
\usepackage{booktabs}

\usepackage[section]{placeins}

\begin{document}

%%%%%%%%%%%%%%%%%%%%%%%%%%%%%%%%
\title{Hawking-radiation–ignited autocatalytic formation of primordial black holes}
%Hawking-Radiation–Induced Primordial Black Hole Cascades and Formation of Planck-Mass Relics}
%%%%%%%%%%%%%%%%%%%%%%%

%\title{The role of quantum pressure in DM}

\author{Alexander Yakimenko}
\affiliation{Department of Physics, Taras Shevchenko National University of Kyiv,
64/13, Volodymyrska Street, Kyiv 01601, Ukraine}
\affiliation{Dipartimento di Fisica e Astronomia ’Galileo Galilei’,
Universit{\'a} di Padova, via Marzolo 8, 35131 Padova, Italy}

\begin{abstract}
We propose and analyze an autocatalytic mechanism in which bursts of Hawking
radiation from evaporating micro–primordial black holes (PBHs) trigger the
collapse of near-critical plasma overdensities. In a primordial plasma seeded
with such patches, this feedback self-organizes into a traveling ignition
front that successively forms new PBHs and then self-quenches as the Universe
expands. A minimal reaction–diffusion model yields conservative criteria for
ignition and freeze-out and predicts a stochastic gravitational-wave
background with a sharp causal low-frequency edge set by the freeze-out
correlation length and largely insensitive to Planck-scale PBH endpoint
microphysics. The resulting sub-Hz–to–audio band and amplitudes satisfy
cosmological energy-injection bounds, providing a clean, testable target for
forthcoming gravitational-wave observatories.
\end{abstract}
\maketitle
%\textit{Introduction.}-- 
If dark matter is composed of primordial black holes (PBHs) -- a minimal,
purely gravitational candidate -- it is subject to tight multi–probe
constraints from CMB anisotropies, dynamics, accretion, and lensing
\cite{CarrKohriSendoudaYokoyama2021,Carr2020,CarrKuhnelPLN2022,Khlopov2010}. A
robust low–mass bound follows from Hawking radiation (HR)
\cite{Hawking1975}: a black hole of mass $M$ emits thermally at
$T_H=(T_{\rm Pl}/8\pi)(M_{\rm Pl}/M)$, so its semiclassical lifetime
scales as $\tau\propto M^{3}$; PBHs with
$M\lesssim 10^{14}\,\mathrm{g}$ evaporate by today
\cite{Page1976a,CarrKohriSendoudaYokoyama2010,GreenKavanagh2021,Carr2016},
and $T_H$ approaches the Planck scale for $M\sim M_{\rm Pl}$. Quantum
backreaction and possible discrete line structure are then expected to
modify the emission, making both the spectrum and effective lifetime
sensitive to quantum–gravity microphysics
\cite{ParikhWilczek2000,ArzanoMedvedVagenas2005,BekensteinMukhanov1995}.
A widely discussed possibility is that evaporation terminates in a
stable Planck–mass relic (PMR)
\cite{Chen2005PlanckRemnants,Carr2020,CarrKuhnelPLN2022,Maldacena2021MagneticBH,
Pacheco2018UniversePRBH,DomenechSasaki2023,DaviesEassonLevin2025NSBHDM,
Calza2025PRD_I,Calza2025PRD_II}, with several independent theoretical
motives for such stabilized endpoints
\cite{Bekenstein1974,BekensteinMukhanov1995,AdlerChenSantiago2001,
Hayward2006PRL,Dymnikova1992,MazurMottola2004PNAS,HoldomRen2017,
AydemirDonoghuePiazza2020,RovelliVidotto2014PRD}. Nonemitting relics are compatible with current constraints
\cite{MacGibbon1987,Barrau2019,Calmet2021,Carr2020,DomenechSasaki2023,
AcharyaKhatri2020,Planck2018}, but the key challenge for PMR dark
matter is achieving a sufficient relic abundance without violating these bounds. Standard radiation–era
collapse ties PBH masses to the horizon at formation; critical collapse
broadens but does not break that link
\cite{ShibataSasaki1999,Harada:2013epa,Musco2019,NiemeyerJedamzik1998PRL},
while other proposed micro–PBH channels invoke enhanced small–scale
power, topological defects, or first–order phase transitions
\cite{ClesseGarciaBellido2015, Clesse2018SevenHints, CarrKohriSendoudaYokoyama2021,Carr2020,GreenKavanagh2021,Sasaki2018}.
By contrast, we introduce a subhorizon, HR–driven
ignition mechanism that seeds micro–PBHs without extreme fine–tuning of
the primordial initial conditions.

We propose that Hawking bursts from evaporating micro–PBHs can stimulate
further PBH formation in the early Universe. A seed population of micro–PBHs ($M_{\rm Pl}<M\ll M_\odot$) emits short EM/hadronic bursts whose small mean
free paths at early times enable rapid, localized energy deposition in
near–threshold regions, transiently reducing pressure support and
pushing some patches above the collapse threshold. Because such
near–critical regions are parametrically more abundant than the
spontaneously collapsing tail, modest localized deposits can
substantially enhance the formation rate without extreme tuning of
primordial fluctuations \cite{NiemeyerJedamzik1998PRL,Musco2019}. In
this PBH–mediated autocatalytic regime, the plasma is the fuel, HR acts as the catalyst, and micro–PBHs are the intermediates:
newborn micro–PBHs evaporate promptly, their bursts drive further
collapses, and the net multiplication is briefly supercritical until
expansion and transport weaken deposition, leaving a saturated relic
abundance consistent with BBN and CMB energy–injection limits
\cite{AcharyaKhatri2020,Planck2018}. At the collective level, this
ignition admits a minimal reaction–diffusion description of
energy–limited conversion of plasma into micro–PBHs; the resulting
traveling front forms new PBHs, self–quenches as the Universe expands,
and fixes a freeze–out scale that imprints a sharp causal low–frequency
edge and a finite, nearly flat plateau in the stochastic
gravitational–wave spectrum, largely independent of endpoint
microphysics.

%%%%%%%%
\paragraph*{Minimal kinetic model for the ignition phase.--}
%%%%%%%%%%
We focus on a short interval $t_0\le t\ll H^{-1}$ when the plasma is
still dense and contains near–threshold patches that can form new PBHs.
On this timescale we work in a small comoving region where expansion is
negligible and the plasma is approximately homogeneous.

Let $\rho_r(t)$ be the local plasma energy density available for
compaction growth within this control volume (the ignition–accessible
reservoir), and let $\rho_{\rm PBH}(t)$ be the energy density in active
micro–PBHs (those currently emitting Hawking bursts). Their coarse–grained
evolution is
\begin{equation}
\dot{\rho}_{\rm PBH}
= -\gamma\,\rho_{\rm PBH}+\alpha\,\rho_{\rm PBH}\,\rho_r
\;+\;S(\rho_r,t).
\label{eq:dn_dt}
\end{equation}
The first term describes evaporation, with an effective loss rate
$\gamma$ that returns energy to the plasma via HR bursts. The second
term, $\alpha\,\rho_{\rm PBH}\rho_r$, describes HR–catalyzed conversion
of plasma into new micro–PBHs; $\alpha$ encodes short–mfp EM/hadronic
deposition, response time, and near–threshold statistics, and does not
describe accretion–driven growth of existing PBHs. The source
$S(\rho_r,t)$ accounts for spontaneous (non–catalyzed) PBH production
from rare supercritical overdensities; we model it as
$S(\rho_r)=\beta\,\rho_r^2$, where $\beta$ parametrizes the small tail
probability in a Gaussian radiation field and is taken constant over the
short ignition interval. If a preexisting horizon–scale PBH population
is assumed, one may instead set $S=0$ during ignition and encode seeding
in the initial compact component.

\begin{figure}[t]
%\centering
\includegraphics[width=\linewidth]{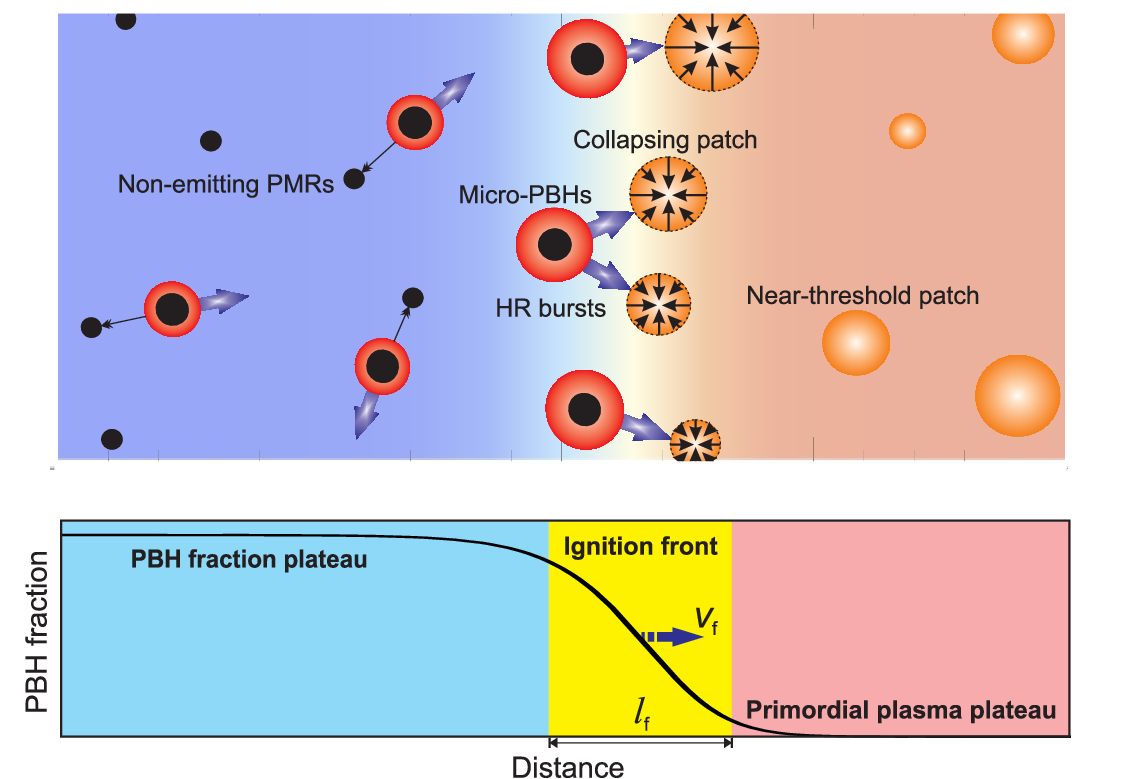}
\caption{{Schematic of a Hawking–radiation–triggered ignition cascade.}
Micro–PBHs emit short HR bursts (blue arrows) that deposit energy within a causal patch (orange shell), tipping nearby near-threshold {patches} to collapse and creating new {micro–PBHs},  that launch further bursts and finally evaporate to non-emitting {PMRs}.
Bottom: rescaled PBH density across the ignition front, which propagates with speed $v_f$ and has thickness $\ell_f$.
(Not to scale.)} \label{fig:Schematics}
\end{figure}
%%%%%%%%%%%%%%%%%%%%%%

During the brief ignition stage we treat a small comoving region as an
active reservoir. HR from PBHs splits into locally deposited
EM/hadronic secondaries, which feed this reservoir, and escaping
channels (neutrinos, gravitons, long–mfp photons) that do not return to
drive collapse; the latter are absorbed into the effective gain and
diffusion coefficients and into the finite ignition duration. Over this
short interval we approximate the active energy as constant,
$\rho_r(t)+\rho_{\rm PBH}(t)\simeq\varepsilon_0$, with only a tiny
fraction of the cosmological radiation bath—near–threshold
patches—participating in ignition, so global depletion is negligible.
Within such a patch we set $\rho_r\simeq\varepsilon_0-\rho_{\rm PBH}$
and reduce the dynamics to a single evolution equation for
$\rho_{\rm PBH}$. In the initial regime
$\rho_{\rm PBH}\ll\rho_r\simeq\varepsilon_0$ a linearization gives
$
\dot{\rho}_{\rm PBH}\simeq(\alpha\,\varepsilon_0-\gamma)\,\rho_{\rm PBH}
+\beta\,\varepsilon_0^{2}.$
Defining $\varepsilon_{\rm th}=\gamma/\alpha$, 
any nonzero seed grows if
$\varepsilon_0>\varepsilon_{\rm th}$ (catalyzed gain outpaces
evaporation), whereas for $\varepsilon_0\le\varepsilon_{\rm th}$ active
PBHs decay and no ignition occurs. As expansion reduces $\rho_r(t)$ the
condition $\lambda=\alpha\,\rho_r(t)-\gamma\le 0$ provides a practical
freeze–out criterion. Non–emitting PMRs—if they exist—appear only after
active micro–PBHs have evaporated and therefore do not affect ignition
kinetics.

%%%%%%%%%%
%\emph{Minimal kinetic model.}
%%%%%%%%%%
%%%%%%%%%%
%\emph{Homogeneous cell.}
%%%%%%%%%%
Normalizing $y=\rho_{\rm PBH}/\varepsilon_0$ gives $\dot y=f_{\rm r}(y)$ with
$
f_{\rm r}(y)=\beta\,\varepsilon_0(1-y)^2+\alpha\,\varepsilon_0(1-y)\,y-\gamma\,y
=(\beta-\alpha)\,\varepsilon_0\,(y-y_1)(y-y_2),
$
where $y$ is the compact fraction and $y_1<y_2$ are the real zeros for
$0\le\beta<\alpha$. When the catalytic gain is subcritical ($\lambda<0$),
quadratic seeding and evaporation balance at a small quasi–steady
baseline
$
y_{*}\simeq{\beta\,\varepsilon_{0}}/({\gamma-\alpha\,\varepsilon_{0}})
$ for $y\ll 1$. For $\lambda>0$ any nonzero seed grows; if
$\alpha>\beta$, the upper root $y_{2}$ is globally attracting on
$0\le y\le 1$ and $y_{1}$ is repelling, whereas for $\alpha\le\beta$
stability is reversed and no self–sustained ignition occurs. Thus
$\lambda$ controls ignition onset and sets freeze–out once dilution
drives $\lambda\to 0$, while the nonlinear reaction self–saturates the
evolution toward its attractive fixed point. We work in the weak–seeding
regime $0<\beta\ll\alpha$, but the conclusions are unchanged if
$\beta=0$ and seeding is purely initial.

The rescaling $u=(y-y_1)/(y_2-y_1)\in(0,1)$ reduces the dynamics to the
logistic form $\dot u = r\,u(1-u)$, with mesoscopic growth rate $r$
setting the saturation timescale $r^{-1}$ and
$
r^2 = {(\alpha\,\varepsilon_0-\gamma)^2
        + 4\,\gamma\,\beta \,\varepsilon_0} \,.
$
Having established local ignition and saturation in a homogeneous cell,
we next ask how spatial inhomogeneities and effective transport
organize ignition into traveling fronts and fix the freeze–out
correlation length, as illustrated in Fig.~\ref{fig:Schematics}.
%%%%%%%%%%
\paragraph*{Spatial transport and Fisher--KPP fronts. -- }
%%%%%%%%%%
To describe how ignition organizes in space we now include spatial
inhomogeneity and transport. Coarse–graining at mesoscopic scales,
$\ell_{\rm mfp}\ll L\ll c/H$ and $t_{\rm dep}\ll t\ll H^{-1}$, makes the
dominant spatial coupling effectively diffusive. Here
$t_{\rm dep}\sim\mathcal{O}(1\!-\!10)\,\ell_{\rm mfp}/c$ is the
EM/hadronic deposition time, with
$t_{\rm dep}\ll t_{\rm sound}=R/c_s\ll H^{-1}$. This diffusion
summarizes two processes distinct from the local reaction rate:
(i) short–mfp EM/hadronic cascades from Hawking bursts redistribute
energy quasi–locally before full thermalization, inducing a nearest–neighbor,
effectively diffusive coupling of the compact fraction (akin to early–Universe
energy–deposition kernels in BBN/CMB analyses \cite{Slatyer2016b}); and
(ii) stochastic recoil from anisotropic Hawking emission drives a random
walk of evaporating micro–PBHs \cite{Page1976a,ArbeyAuffinger2019,
AuffingerArbey2021}, which coarse–grains to an additional diffusion
channel. 
We subsume the effective diffusivity $D$ and the logistic reaction term
$r\,u(1-u)$ into the classical Fisher–KPP equation
\begin{equation}
\partial_t u(x,t) = D\,\partial_x^2 u(x,t)+ r\, u(x,t)\left\{1-u(x,t)\right\},
\label{eq:FisherKPP}
\end{equation}
which captures diffusion–limited autocatalytic growth
\cite{Fisher1937,KPP1937}. This reduction to a single reaction–diffusion
equation is analogous to standard front–propagation treatments in
combustion and population dynamics, where detailed conservation laws are
coarse–grained into an effective Fisher–KPP front once a scale
separation between a thin active zone and a background reservoir is
assumed \cite{Fisher1937,KPP1937,EbertVanSaarloos2000}.

\begin{figure}[t]
\includegraphics[width=1.\linewidth]{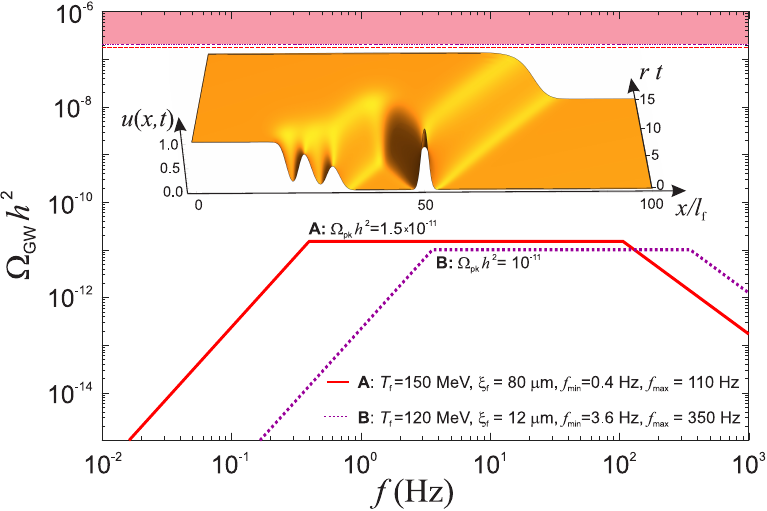}
\caption{Stochastic background from HR–ignited fronts: causal
$f^{3}$ rise, flat plateau of height $h^{2}\Omega_{\rm pk}$, and UV
fall $\propto f^{-2}$. Two benchmark spectra are shown: Case A (solid
red) and Case B (dotted blue). Horizontal lines show the CMB–based
$\Delta N_{\rm eff}$ cap (A: red dashed; B: blue dotted), and the shaded
band marks the excluded region. Inset: Fisher–KPP ignition selects a
pulled traveling front. Starting from a steep profile with localized
bumps, diffusion and growth erase small–scale structure and the front
converges to speed $v_c$ and width $\ell_f$.
}\label{fig:GW}
\end{figure}

Equation~(\ref{eq:FisherKPP}) admits traveling–front solutions
$u(x,t)=F(\xi)$, $\xi=x-vt$, with $F(-\infty)=1$ and $F(+\infty)=0$.
Linearizing the leading edge ($F\ll1$) gives $F\sim e^{-k\xi}$ and
$v(k)=Dk+r/k$, minimized at $k_c=\sqrt{r/D}$, so that
$v_c=2\sqrt{Dr}$ and $\ell_f\sim\sqrt{D/r}$. An analytic solution exists
for $v_{\rm AZ}=\tfrac{5}{2\sqrt{6}}v_c>v_c$ with
$F(\xi)=[1+\exp(\xi\sqrt{r/(6D)})]^{-2}$ \cite{AblowitzZeppetella1979},
while no elementary form is known at $v=v_c$. For ignition–type (steep)
initial data the KPP selection principle yields convergence, up to a
translation, to the minimal wave \cite{KPP1937,AronsonWeinberger1978,
Bramson1983}. Traveling fronts are nonlinearly stable: pushed
($v>v_c$) waves relax exponentially, whereas pulled ($v=v_c$) waves
relax algebraically with the Bramson shift \cite{Sattinger1976,
FifeMcLeod1977,EbertVanSaarloos2000,AveryScheel2021}. To illustrate
selection, the inset of Fig.~\ref{fig:GW} shows a numerical solution of
Eq.~(\ref{eq:FisherKPP}) started from a steep sigmoid profile with
three localized bumps. Small–scale
inhomogeneities are smoothed by diffusion and absorbed into the leading
edge, so the system rapidly forgets the initial details: a pulled front
emerges with asymptotic speed $v\to v_c$ and thickness $\sim\ell_f$.
Hence the freeze–out correlation length is mesoscopic: it is set by
coarse–grained transport and growth parameters and by the short ignition
window, not by the microscopic structure of the seeds.

%%%%%%%%%%
\paragraph*{Ignition viability. --}
%%%%%%%%%%
In the standard picture an evaporating PBH produces an outward,
high–pressure flash that \emph{opposes} collapse in the primordial
plasma; this is true for free–streaming channels (gravitons, neutrinos)
and for spatially homogeneous heating. Here we instead isolate a narrow
causal window in which short–mfp EM/hadronic secondaries from Hawking
bursts are promptly absorbed and locally raise the inner compaction of
near–critical patches. Because collapse thresholds depend on profile as
well as amplitude, a small, well–timed local deposit can sharpen the
core compaction at fixed $\delta = \delta\rho_r/\rho_r$ and lower the
effective threshold \cite{NiemeyerJedamzik1998PRL,Musco2019,Musco2021}. We do
\emph{not} invoke exotic focusing (“BH out of light’’) or ultra–rare
head–on collisions; free–streaming power is treated as noncatalytic,
consistent with quantum focusing/inequality bounds that disfavor horizon
formation from outgoing radiation \cite{BoussoEtAl2016QFC,FordPfenningRoman1998}.
Ignition therefore operates only in a finite PBH mass window: very light
micro–PBHs have intense Hawking power but limited total energy and can
tip only a modest number of near–threshold patches, whereas very massive
PBHs evaporate so slowly that most of their output is released after the
plasma has cooled and no longer satisfies the tipping criterion.
%%%%%
\paragraph*{Quantitative tipping criterion. --}
%%%%%%
Consider a near-critical patch with $\delta=\delta_c-\varepsilon$ ($0<\varepsilon\ll1$) at horizon entry. A nearby micro–PBH releases a terminal burst with energy $\Delta E_{\rm burst}$. The short–mfp electromagnetic/hadronic component deposits a temperature–dependent fraction $f_{\rm dep}(T)$ locally within a causal volume $V_{\rm dep}=4\pi R^3/3$ over a sub-acoustic time $t_{\rm dep}\ll t_{\rm sound}\sim R/c_s$ ($c_s\simeq c/\sqrt{3}$), so pressure cannot respond. The resulting compact load $\chi= f_{\rm dep}\Delta E_{\rm burst}/(\varepsilon\,\rho_r\,V_{\rm dep})$ must satisfy $\chi\gtrsim 1$ to tip the patch; i.e., the locally deposited energy must exceed the deficit $\varepsilon\,\rho_r V_{\rm dep}$ required to reach threshold. Equivalently, in compaction variables the condition reads $\Delta\mathcal{C}(R)=(2G/Rc^{4})\,\Delta E_{\rm dep}\gtrsim\Delta\mathcal{C}_{\rm req}$ with $\Delta\mathcal{C}_{\rm req}= \mathcal{C}_c-\mathcal{C}_{\max}(\delta)\simeq(\partial\mathcal{C}_{\max}/\partial\delta)_{\delta_c}\,(\delta_c-\delta)=\kappa\,\varepsilon$, where $\mathcal{C}_{\max}$ is the peak Misner–Sharp compaction and $\kappa=\mathcal{O}(1)$ encodes the weak profile/EOS dependence near threshold \cite{Harada:2013epa,Musco2019,NiemeyerJedamzik1998PRL}. 
Here $\Delta E_{\rm dep}=f_{\rm dep}\,\Delta E_{\rm burst}$ and $R\lesssim R_{\rm comp}$ is the radius that maximizes $\mathcal{C}$. Only compact, rapid deposition within $R_{\rm comp}$ increases $\mathcal{C}$; short–mfp EM/hadronic channels contribute to $f_{\rm dep}(T)$, whereas free–streaming species (neutrinos, gravitons) deposit negligibly on sub–acoustic times and thus do not tip near–critical patches~\cite{Slatyer2016b,Poulin2017,BashinskySeljak2004}. Using $R\sim c_s t_{\rm dep}$ and $\rho_r\propto T^4$ implies $\chi\propto f_{\rm dep}\Delta E_{\rm burst}\,T^{-4}\,t_{\rm dep}^{-3}$, i.e., earlier epochs require faster, more local deposition, while shorter $t_{\rm dep}$ significantly increases the tipping effectiveness.

%\textit{Target abundance and %cumulative loading.—}
Quantitatively, the target population is substantial even though PBH
formation must remain rare. For a Gaussian field with variance
$\sigma^2\ll\delta_c^2$, the ratio of near–threshold patches in the
shell $[\delta_c-\varepsilon,\delta_c]$ to spontaneously supercritical
regions $(\delta>\delta_c)$ at the same smoothing scale is
$
\mathcal{R} = {P_{\rm sub}(\varepsilon)}/{P_{\rm tail}}
\simeq {\varepsilon\,\delta_c}/{\sigma^2}.
$
With $\delta_c\in[0.4,0.5]$, $\varepsilon\in[0.02,0.08]$, and
$\sigma\in[0.03,0.07]$ one finds $1.6\le\mathcal{R}\le 44$, so tip–able
patches can outnumber spontaneous collapses by factors of a few to tens.

Several effects further aid threshold crossing: multiple subcritical
deposits can accumulate within one acoustic time before pressure
rebound, centrally concentrated loading lowers the effective collapse
threshold relative to uniform heating
\cite{NiemeyerJedamzik1998PRL,Musco2019}, and near the QCD crossover the
temporary reduction of $c_s^2$ and enhanced bulk viscosity promote
compaction growth during the first compression \cite{Karsch2008,Meyer2007}.

Finally, because a PMR is extremely heavy
($M_{\rm PMR}\simeq M_{\rm Pl}$), its number density is tiny:
$n_{{\rm PMR},0}\simeq \rho_{\rm DM}/M_{\rm PMR} \sim 10^{-25}\,{\rm cm^{-3}}$
(cosmic mean; spacing $\sim 2000~{\rm km}$). Even in the local halo with
$\rho_{\rm DM}\sim 0.3\,{\rm GeV\,cm^{-3}}$ one finds
$n_{\rm PMR}\sim 2.5\times10^{-20}\,{\rm cm^{-3}}$ (spacing
$\sim 30~{\rm km}$). PMR production is therefore energy– rather than
multiplicity–limited: what matters is the total energy converted into
relics, so only a tiny global fraction of the plasma need pass through
PMRs during ignition, without requiring many successful collapses per
causal region.

%%%%%%%%%%%%%%%%%%%%%%
\paragraph*{SGWB channels during ignition.—}
%%%%%%%%%%%%%%
The ignition process acts as a broadband GW source by driving
overpressured regions whose bulk motion generates time–dependent
anisotropic stress. The resulting stochastic background can receive
power from expanding fronts around igniting PBHs, acoustic post–shock
sound waves, and a later turbulent tail
\cite{HuberKonstandin2008,Hindmarsh2017,CapriniFigueroa2018,Caprini2009};
ignition–induced inhomogeneities can also source a second–order
(scalar–induced) background \cite{KohriTerada2018,DomenechSasaki2023},
while direct Hawking gravitons yield an ultra–high–frequency, nearly
thermal component \cite{MacGibbon1991}. In what follows we focus on the
lower–frequency environmental signal from ignition–driven flows, whose
edge–plus–plateau structure is set by the freeze–out correlation length.

The ignition stage excites mesoscopic bulk flows during a brief active
interval and then shuts off. The ignition stage excites a Fisher–KPP front of width $\ell_f$ and
speed $v_c$ (see Fig.~\ref{fig:GW}),  so propagation during the active
interval $\Delta t_{\rm ign}$ leaves a freeze–out correlation length
$\xi_f\simeq\max(\ell_f,\,v_c\Delta t_{\rm ign})$ and a source
coherence scale $\ell_{\rm cut}=\max(\ell_f,\,\ell_{\rm micro})$,
where $\ell_{\rm micro}$ is the short–mean–free–path EM/hadronic
deposition footprint and $c_s$ is the sound speed; the associated
coherence time satisfies $\tau_{\rm coh}\sim\ell_{\rm cut}/c_s$.

Defining the redshift factor $\zeta_f = a_f/a_0$, a finite correlation
length $\xi_f$ fixes the causal low–frequency edge
$f_{\min,0} = \zeta_f c/(2\pi\xi_f)$ and implies the standard causal
rise $\Omega_{\rm GW}\propto f^3$ for $f\lesssim f_{\min,0}$
\cite{CapriniFigueroa2018,DurrerCaprini2003}. Here
$\Omega_{\rm GW}(f) = \rho_c^{-1}\,d\rho_{\rm GW}/d\ln f$ with
$\rho_c = 3H_0^2/(8\pi G)$ the critical density. For frequencies
$f_{\min,0}\lesssim f\lesssim f_{\max,0}$, with
$f_{\max,0} = \zeta_f c_s/(2\pi\ell_{\rm cut})$, the source is
short–correlated in time (correlation time $\tau_{\rm coh}$), so its
unequal–time correlator is effectively white over the band of interest.
Spatial correlations decay over $\xi_f$, and for modes
$k\lesssim\xi_f^{-1}$ the equal–time power spectrum of the anisotropic
stress is approximately $k$–independent, so different correlation
volumes add incoherently. The convolution with the GW Green’s function
then produces a broad maximum with weak scale dependence, which we
approximate as a flat plateau
$\Omega_{\rm GW}(f)\simeq \Omega_{\rm pk}$ for
$f_{\min,0}\lesssim f\lesssim f_{\max,0}$. At higher frequencies
$f\gtrsim f_{\max,0}$ the finite coherence time and length suppress the
source on small scales, leading to a decaying UV tail that we model as
$\Omega_{\rm GW}\propto f^{-p}$ with $p$ a free index; values
$p\gtrsim 2$ are motivated by analytic estimates and simulations of
short–lived, causal sources such as phase transitions, turbulence, and
primordial magnetic fields
\cite{Caprini:2009fx,Caprini:2001nb,Caprini2009,CapriniFigueroa2018,
Hindmarsh2017,Weir2018}.

The two spectra in Fig.~\ref{fig:GW} are drawn from the interior of a
conservative mesoscopic domain defined by
$\ell_f\ge \ell_{\rm micro}$, subsonic propagation $v_c\le c_s$, and
strong local absorption $D\ll c_s\ell_{\rm micro}$. These choices ensure
that the ignition front operates within the microphysical window
identified earlier from the tipping parameter $\chi\gtrsim 1$, with $r$
and $D$ large enough to form a front on timescales $\ll H^{-1}$ yet
small enough to remain compatible with cosmological energy–injection
bounds.
In both benchmarks
$\ell_f>\ell_{\rm micro}$, so the coherence scale is set by the front,
$\ell_{\rm cut}=\ell_f$, and $f_{\min,0}$ and $f_{\max,0}$ follow
directly from $\xi_f=v_c\Delta t_{\rm ign}$ and $\ell_f$ as above.
Thus, for given $(\xi_f,\ell_f)$ (as in Fig.~\ref{fig:GW}) the
broken–power–law template is fixed up to the plateau height
$\Omega_{\rm pk}$ and the UV index $p$.

For each benchmark we adopt a conservative plateau normalization
$h^{2}\Omega_{\rm pk}$ that encodes the (unknown) GW conversion
efficiency of the ignition stage via
$\Omega_{\rm pk}\sim\epsilon_{\rm GW}\,\Omega_{\rm kin}$, where
$\Omega_{\rm kin}$ is the bulk–flow energy fraction. A first–principles
normalization would require fully coupled relativistic
radiation–hydrodynamics with kinetic transport of EM/hadronic cascades
and Planck–scale endpoint physics, which lies beyond this
proof–of–concept study. Instead, we calibrate $\epsilon_{\rm GW}$ using
a well–studied class of short–lived relativistic fluid sources—sound
waves from first–order phase transitions—where simulations and analytic
templates yield $h^{2}\Omega_{\rm pk}\sim10^{-12}$–$10^{-9}$ for
temporally short, subsonic sources
\cite{Hindmarsh2014PRL,Hindmarsh2017,Caprini2016PT,Weir2018}. Our
benchmark $h^{2}\Omega_{\rm pk}\sim10^{-11}$ is therefore conservative
for a generic short–lasting plasma source with similar correlation
scales and remains well below the $\Delta N_{\rm eff}$ cap.

We translate the standard extra–radiation constraint into a GW integral
bound using the relation between $\Delta N_{\rm eff}$ and the integrated
GW density \cite{CapriniFigueroa2018,Planck2018},
$\int d\ln f\,h^2\Omega_{\rm GW}(f)\le 5.6\times 10^{-6}\,\Delta
N_{\rm eff}$, where $h = H_0/(100\,{\rm km\,s^{-1}\,Mpc^{-1}})$. For
the broken–power–law template above this integral evaluates to
$h^2\Omega_{\rm pk}\,I$ with
$I = 1/3 + \ln(f_{\max,0}/f_{\min,0}) + 1/p$, so the plateau height is
capped at
$h^2\Omega_{\rm pk}^{\max}=5.6\times 10^{-6}\Delta N_{\rm eff}^{\max}/I$.
In our benchmarks we adopt $p=2$ when evaluating $I$, consistent with UV
slopes found for short–lived, causal sources and conservative for the
$\Delta N_{\rm eff}$ bound: for fixed $\Omega_{\rm pk}$, larger $p$
would only decrease $I$ and thus weaken the constraint. In
Fig.~\ref{fig:GW} we draw, for each benchmark, a horizontal cap at
$h^2\Omega_{\rm pk}^{\max}$ (we use a CMB–based
$\Delta N_{\rm eff}^{\max}=0.2$). This cap applies only to the decoupled GW component. At the epoch of interest,
$T\sim 50$–$150~\mathrm{MeV}$, neutrinos remain tightly coupled on Hubble
times and thus do not yet contribute to $\Delta N_{\rm eff}$. If ignition
persisted past neutrino decoupling, any late power injected into
free–streaming neutrinos would enter $\Delta N_{\rm eff}$ and further tighten
the radiation budget.

%%%%%%%%%
\paragraph*{Conclusions.--}
%%%%%%%%%%
Hawking–radiation–driven ignition can operate briefly but efficiently in
the radiation era, amplifying near–critical collapses while the plasma
is dense and then self–quenching as dilution shuts off the gain. This
scenario predicts a robust stochastic gravitational–wave background with
a sharp causal low–frequency edge set by the freeze–out correlation
length—the primary observable—largely insensitive to the unknown
Planck–scale evaporation microphysics, so that the edge–plus–plateau
spectrum persists even if all micro–PBHs fully evaporate. If evaporation
halts at the Planck scale, the same ignition epoch could also yield a
population of Planck–mass relics as a purely gravitational dark–matter
candidate. Thus the ignition mechanism provides an independently
testable route to PBH formation, and its GW signature offers a rare
mesoscopic window on early–Universe microphysics.

%%%%%%%%%%%%%%%%%%%
\paragraph*{Acknowledgments. --}
Valuable discussions with Oleksandr Sobol and Luca Salasnich are gratefully acknowledged.

\bibliographystyle{apsrev4-2}
\bibliography{Refs}
%\newpage

%\input{Supplemental material}

\end{document}